\documentclass[useAMS,usenatbib,letter]{mn2e}
\usepackage{psfig} 
\usepackage[dvips]{graphicx}
\usepackage{natbib}
\def\mnras{MNRAS}
\def\apj{ApJ}
\def\aap{A\&A}
\def\apjl{ApJL}
 
\def\apjs{ApJS}

\def\pasp{PASP}

\def\xmm{{\sl XMM-Newton}}
\def\xte{{\sl RXTE}}
\def\ngc{{NGC~3783}}

\title[X-ray/optical Correlation and delays in \ngc]{Correlation and time delays of the X-ray and optical emission of the Seyfert Galaxy \ngc}  
\author[P. Ar\'evalo et al.]{P. Ar\'evalo$^{1,2}$\thanks{E-mail: patricia@shao.ac.cn}, P. Uttley$^{3}$, P. Lira$^{4}$, E. Breedt$^{3}$, I. M. McHardy$^{3}$, E. Churazov$^{5,6}$ \\ 
$^1$Shanghai Astronomical Observatory, 80 Nandan Road, Shanghai 200030, China\\
$^2$Departamento de Ciencias Fisicas, Universidad Andres Bello, Av.
Republica 252, Santiago, Chile\\
$^3$School of Physics and Astronomy, University of Southampton, Southampton SO17 1BJ, UK\\
$^4$Departamento de Astronom\'ia, Universidad de Chile, Casilla 36-D, Santiago, Chile\\
$^5$Max-Planck-Institut f\"ur Astrophysik, Karl-Schwarzschild-Strasse 1, 85741 Garching, Germany\\
$^6$Space Research Institute (IKI), Russian Academy of Sciences, Profsoyuznaya 84/32, 117997 Moscow, Russia\\
}

\begin{document}
\date{Received /Accepted}
\pagerange{\pageref{firstpage}--\pageref{lastpage}} \pubyear{2008}

\maketitle
\label{firstpage}
 
\begin{abstract}
We present simultaneous X-ray and optical B and V band light curves of the Seyfert Galaxy \ngc\ spanning 2 years. The flux in all bands is highly variable and the fluctuations are significantly correlated. As shown before by Stirpe et al.\ the optical bands vary simultaneously, with a delay of less than 1.5 days but both B and V bands lag the X-ray fluctuations by 3--9 days. This delay points at optical variability produced by X-ray reprocessing and the value of the lag places the reprocessor close to the broad line region. A power spectrum analysis of the light curve, however, shows that the X-ray variability has a power law shape bending to a steeper slope at a time-scale $\sim 2.9$ days while the variability amplitude in the optical bands continues to grow towards the longest time-scale covered, $\sim 300$ days. We show that the power spectra together with the small value of the time delay is inconsistent with a picture where \emph{all} the optical variability is produced by X-ray reprocessing, though the small amplitude, rapid optical fluctuations might be produced in this way. We detect larger variability amplitudes on long time-scales in the optical bands than in the X-rays. This behaviour adds to similar results recently obtained for at least three other AGN and indicates a separate source of long term optical variability, possibly accretion rate or thermal fluctuations in the optically emitting accretion disc. 

\end{abstract}

\section{Introduction}
The optical continuum emission in Active Galactic Nuclei (AGN) almost certainly originates as thermal emission from the accretion disc \citep{koratkarblaes99} but the origin of the X-ray emission in these objects is less clear. The accretion disc is expected to be too cool to produce X-rays and the spectrum in this band is non-thermal, so the existence of a `corona' has been suggested to produce X-ray emission by Compton up-scattering optical/UV photons to higher energies. As it has not been established how this corona is formed, its location and relation to the accretion disc are uncertain, although variability and energetics arguments place it in the innermost regions of the AGN. One way to probe the connection between the accretion disc and corona is to track the variability in X-ray and optical bands simultaneously, to determine their degree of correlation and the relative location of their emitting regions by measuring delays between the fluctuations in both bands.

If the X-rays are produced by a corona in the vicinity of an optically thick accretion disc, it is expected that at least some of the X-ray flux will be intercepted by the disc and be reprocessed thermally into optical emission. The short variability time-scales observed in optical bands and the short delays with respect to the X-ray variations suggest that the optical variability arises from reprocessing of the highly variable X-ray flux \citep{kroliketal91}. The radial temperature profile for a standard accretion disc, $T(R) \propto R^{-3/4}$, places the main optical emitting  regions at different radii in the disc, leading to longer light travel times for longer wavelength optical emission. The wavelength dependent delays predicted by this model are in agreement with the observations of several AGN \citep{wandersetal97,collieretal01,cackettetal07}.

There is an increasing number of objects, however, whose X-ray/optical behaviour is in conflict with a model where reprocessing is responsible for all the optical variability. Long time-scale optical fluctuations have higher amplitude than their X-ray counterparts in at least three cases, NGC~5548 \citep{uttleyetal03}, MR~2251-178 \citep{MR2251} and Mkn~79 \citep{mkn79}. X-rays have been seen to lag the optical bands in e.g.\ NGC~4051 \citep{shemmeretal03} and Mkn~509 \citep{marshalletal08} and \citet{reprocessing} have modeled in detail the reprocessing in the case of NGC~3516 and shown that this process is unable to explain its optical variability. Reprocessing is also challenged by energetics arguments in some AGN, where the luminosity in the optical bands largely exceeds the luminosity in X-rays, making it impossible for the X-rays alone to produce the observed large amplitude optical flux variations \citep{Gaskell07}.   

In \citet{MR2251} we proposed a composite model to explain the X-ray optical correlation in MR~2251-178. We allowed a small fraction of the optical emission to arise from reprocessing, in order to produce the observed small amplitude, rapid optical fluctuations, with a negligible delay with respect to the X-rays. At the same time, the long term variability in both optical and X-ray bands would be modulated by accretion rate fluctuations, allowing different amplitudes of variability in both bands on long and short time-scales. Predictions of this two-component scenario should be tested on AGN of different masses and accretion rates, $\dot m$, since the location of the optical emitting region on a standard accretion disc depends on these parameters \citep[e.g][]{trevesetal88}. Low $\dot m$ and high mass black holes should have a relatively cool disc, placing the optical emitting region at small radii, probably close to the X-ray emitting corona. Therefore, in these cases we should expect a stronger X-ray/optical correlation and variability on similar time-scales. The opposite would happen for low mass and high accretion rate AGN. We are currently building a sample of AGN monitored simultaneously in X-rays and optical bands, which covers a large range in mass and accretion rate to establish the connection between their optical and X-ray emitting regions. Here we present results on our third target, \ngc , which has a black hole mass of $3\times 10^{7} M_\odot$ \citep{revmap} and average accretion rate $\dot m=7\%$ of the Eddington value \citep{Woo}.   

The structure of the AGN in \ngc\ has been studied through several monitoring campaigns. \citet{3783_IRvar} monitored \ngc\ in the optical and infrared bands finding an $\sim 80$ day lag between V and K bands from a 15 year long light curve. This lag is interpreted as the light travel time between the central engine and the hot dust which reprocesses the optical/UV emission and is confirmed and studied in detail by Lira et al.\ (in prep.). The measured lag is at precisely the value expected for the light travel time to the dust at the sublimation radius given the luminosity of \ngc , \citep[see][]{magnum}. The sparse sampling of the long light curves of \citet{3783_IRvar}, however, does not allow a measurement of lags between optical bands and the lack of X-ray coverage precludes a study of the origin of the optical variability. The structure of the broad line region (BLR) has been studied through reverberation mapping campaigns lasting over 7 months, reported in \citet{Reichert} and \citet{Stirpe}. These authors found almost simultaneous variability in the UV and optical continua with amplitude increasing with decreasing wavelength. They also find an $\sim8$ day delay between the continuum and broad-line emission variations, establishing the distance between the BLR and the central continuum source.  The optical/X-ray correlation in \ngc\ on short timescales was investigated by \citet{smith} from simultaneous X-ray and UV observations obtained with \xmm . The $\sim 5$ days long light curves in both bands show significant variability but the data are not sufficient to detect a statistically significant correlation or lag. In this paper we study the optical/X-ray correlation on time-scales of days to hundreds of days using 2-year long, regularly sampled light curves with the aim of establishing the relation between optical and X-ray emitting regions. The monitoring campaigns and data reduction are described in Sec.~\ref{data}. We compare the variability properties of optical and X-ray light curves through the power spectrum in Sec.~\ref{psd} and calculate their cross correlation and time lags in Sec.~\ref{ccf}. We interpret the variability properties and lag time-scales in terms of physical models in Sec.~\ref{discussion} and summarise our conclusions in Sec.~\ref{conclusion}.

\section{Data}
\label{data}
\begin{center}
\begin{figure*}
\psfig{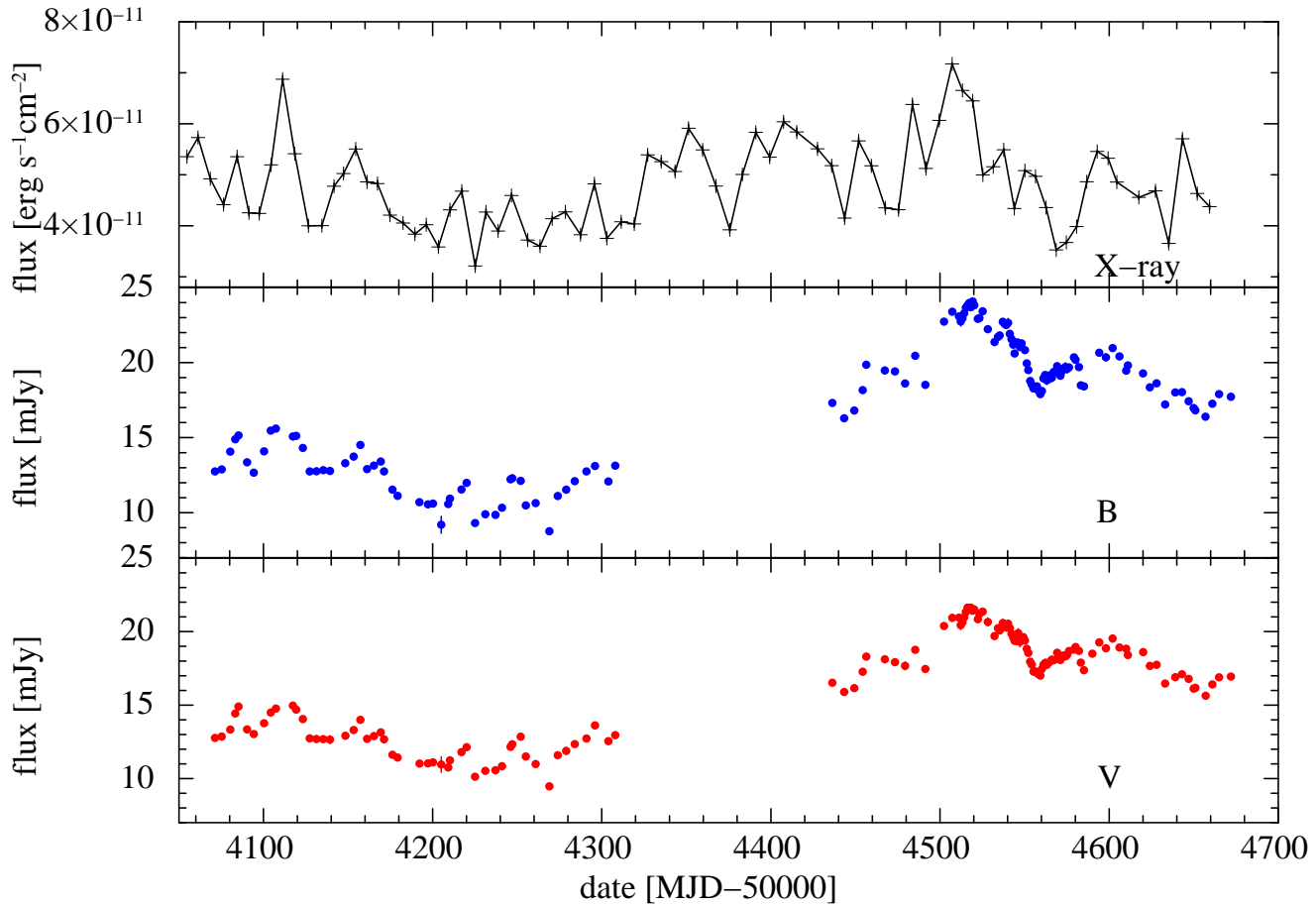}
\caption{\ngc\ light curves. From top to bottom: 2--10 keV X-rays, SMARTS data in the B band and V band. For clarity, the X-ray light curve has been binned in 6 day bins in this plot. The B and V fluxes were measured using psf fitting to the central luminosity peak of the galaxy images so they include some starlight contamination from the underlying galaxy. The intensive sampling in the optical light curves occurs at MJD-50000=4505--4580.}
\label{lcs}
\end{figure*}
\end{center}

We have monitored \ngc\ in the X-ray band using the {\it Rossi X-ray Timing Explorer} \xte\ and in the B and V optical bands with the 1.3m SMARTS telescope in Chile. The resulting light curves are shown in Fig.~\ref{lcs}. Below we give a brief description of each observational campaign.

\subsection{X-ray monitoring with \xte}
\ngc\ was monitored by \xte\ by taking approximately 1~ks exposure
snapshots at regular intervals. For the purposes of the present paper,
we include X-ray data from 2006 November 1 to 2008 August 31. The
sampling pattern varied during this period: exposures were taken every
two days from the beginning of the campaign to 2007 June 29 then every
four days until 2008 February 8. This was followed by an intensive
sampling campaign when observations where made three times per day from
2008 February 8 to 2008 June 3, after this date the sampling returned
to one observation every four days. The data were reduced as
described in \cite{MR2251} and the light curve construction followed
the procedure detailed in the same paper. In particular, the 2--10 keV
flux was estimated by fitting an absorbed power law model to each
snapshot spectrum, where the absorption column was fixed at the
Galactic value, $N_{\rm H}=8.7\times10^{20}{\rm cm}^{-2}$ \citep{Lockman}. The
long-term light curve, binned in 6 day bins is shown in the top panel
in Fig.~\ref{lcs}.

\subsection{B and V band monitoring with SMARTS}

The optical monitoring was performed using the ANDICAM instrument
mounted on the 1.3m SMARTS telescope in Cerro Tololo, Chile. Two B and
V band exposures of \ngc\ were taken for each filter. Observations
were made every four days between 2006 December 1 and 2008 August 31
including a daily sampling period between 2008 February 15 and April
28. The period of daily sampling in the optical was contained within the period of intensive X-ray monitoring.

\begin{center}
\begin{figure}
\psfig{file=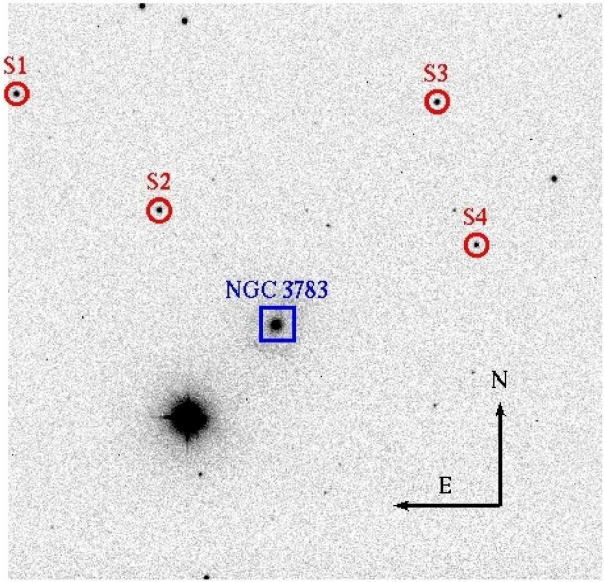,angle=0,width=8cm}
\caption{B band ANDICAM image taken with the 1.3m SMARTS telescope. \ngc\ and the stars used as reference (S1 to S4) are marked and labelled. The reference stars are not saturated and appear in all the campaign images. The N and E arrows are $1\arcmin$ long.}
\label{image}
\end{figure}
\end{center}

We performed psf photometry on \ngc\ and four non-variable stars in
the field of view using the {\sc IRAF} task psf. We constructed
relative flux light curves by dividing the flux measured for \ngc\
nucleus by the sum of the reference star fluxes. The errors were
calculated by propagating the magnitude error produced by the psf
task. We confirmed that the relative flux of the reference stars was
constant throughout the campaign. A B band image of \ngc\ and the
reference stars is shown in Fig.~\ref{image}. We note that the
underlying galaxy flux within the nuclear PSF is not subtracted from
the AGN flux, so a relatively small amount of starlight contamination
remains.

We performed aperture photometry on the reference stars and used the
zero-point correction factors calculated for the SMARTS telescope
every photometric night, together with the airmass of our
observations, to calibrate their magnitudes. We converted these
magnitudes into flux densities at 5500 \AA\ for the V band and
4400\AA\ for B, assuming a flat spectrum ($\alpha=0$) within each band
and used the total flux of the four stars to calibrate the
relative-flux light curve of \ngc . The final light curve fluxes were
corrected for foreground Galactic extinction using $A_\lambda=0.514$ for B
and $A_\lambda=0.395$ for V \citep{Schlegel}. \citet{Winkleretal92}
used several methods to estimate the extinction towards \ngc\ and
concluded that it can be largely or completely accounted for by the
foreground Galactic value, we therefore made no further correction to
the fluxes.

The calibrated B and V band light curves are shown in the middle and
bottom panels in Fig.~\ref{lcs}, respectively. The psf photometry
method was used to include all the point source emission from the
nucleus of the AGN while minimising the contribution from the host galaxy
but it is likely that some contamination remains. We note however that
in the B band light curve the flux varies from under 9 mJy to over 24
mJy, so the majority of the flux measured must be nuclear (assuming
that the star light is constant). In the V band the maximum to minimum
variation is slightly smaller but still more than a factor of 2. This
difference in variability amplitude can be caused by stronger star
light contamination in the V band but it is also possible that the AGN V
emission is intrinsically less variable.

\section {Power Spectrum}
\label{psd}
\begin{center}
\begin{figure}
\psfig{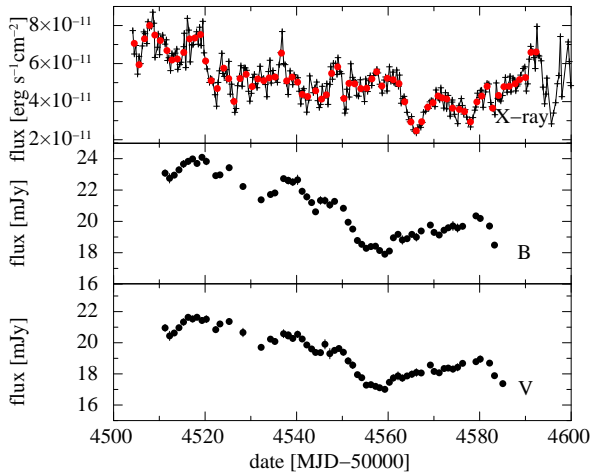}
\caption{Intensive sampling \ngc\ light curves. From top to bottom: 2--10 keV X-rays, SMARTS data in the B band and V band. The top panel shows the 3-times daily sampled light curve in crosses and these data binned in 1 day bins in circles.}
\label{intensive}
\end{figure}
\end{center}

The power density spectrum (PDS) can be used to quantify the variability amplitude as a function of the time-scale of the variations, or correspondingly, of their Fourier frequency. The PDS is constructed through the modulus squared of the discrete Fourier Transform (DFT) \citep{Press}. For the normalisation used in our calculations, the integral of the PDS over frequency equals the normalised variance of the light curve.

\begin{figure}
\psfig{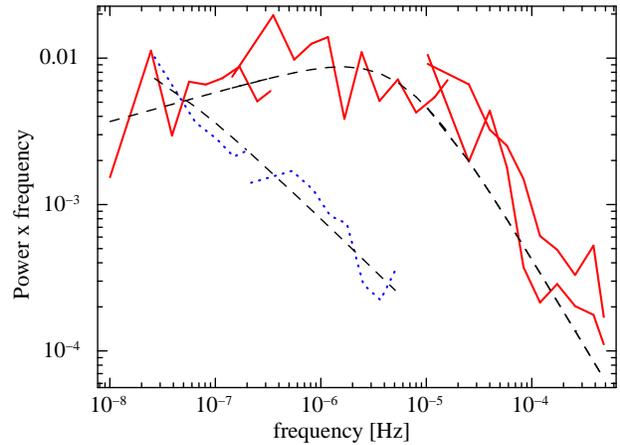}
\caption{Power spectra of \ngc\ in the X-ray band (solid lines) and the optical B band (dotted lines). The dashed lines represent the best-fitting bending power model for each band. The apparent QPO in the X-ray light curve at $f\sim5\times10^{-6}$ Hz is not confirmed by the new intensively sampled data. The X-ray band shows clearly more variability power at high frequencies than the B band but at low frequencies ($\sim5\times10^{-8}$ Hz) their power spectra intersect.}
\label{pds}
\end{figure}

For most AGN X-ray light curves, the PDS has a power law shape
of slope $\sim -1$ bending to a steeper slope at high frequencies
(e.g. Summons et al.\ in prep., \citealt{McHardy4051,McHardyMCG}), which is similar to the PDS found in
stellar mass black hole binaries in the soft state \citep[see][for a review]{uttleyreview}. It is customary
to multiply the variability power by frequency when plotting the PDS,
to highlight deviations of the power law slope from --1 as, in this
case, the low frequency part of the PDS appears approximately flat and
the breaks are more noticeable. We use this standard of presentation in
Fig.~\ref{pds}.

In \citet{summons} we show the X-ray PDS of \ngc\ and explore the
significance of an apparent quasi-periodic oscillation (QPO) at a
frequency of $\sim5 \times 10 ^{-6}$Hz. The new, intensively sampled
light curve obtained for this object, shown in Fig.~\ref{intensive}, with a
sampling rate of three times daily and a length of four months,
covers the frequency range $10^{-7}-3.4\times 10^{-5}$ Hz. This range 
covers the  time-scales corresponding to the peak frequency of the
possible QPO very well and allows us to test its significance
conclusively. We calculated the PDS using the long term \xte\ light
curves and short term \xmm\ light curves discussed in
\citet{summons} and added the new intensive \xte\ data. The
resulting PDS is shown in solid lines in Fig.~\ref{pds}, where the
segments correspond to the different X-ray light curves used.

We fitted a bending power law model defined as
\begin{equation}
P(f)=\frac{Af^{-\alpha_L}}{1+(f/f_b)^{{\alpha_H-\alpha_L}}}
\end{equation}
to the PDS using the Monte Carlo fitting technique {\sc psresp} of \citet{psresp}. The low-frequency slope $ \alpha_L$, the high frequency slope $\alpha_H$, the
bend frequency $f_b$ and the normalisation $A$ were allowed to vary. The best-fitting parameters are $\alpha_L=0.8$, $\alpha_H=2.2$, $f_b=5.8\times 10^{-6}$ Hz, consistent with the values obtained by \citet{summons} for the same bending power law model.  The corresponding model is shown by the dashed line in Fig.~\ref{pds}. The simple bending power law provided an excellent fit to the new data (90\% acceptance probability), making the possible QPO feature unnecessary. 

\begin{figure}
\psfig{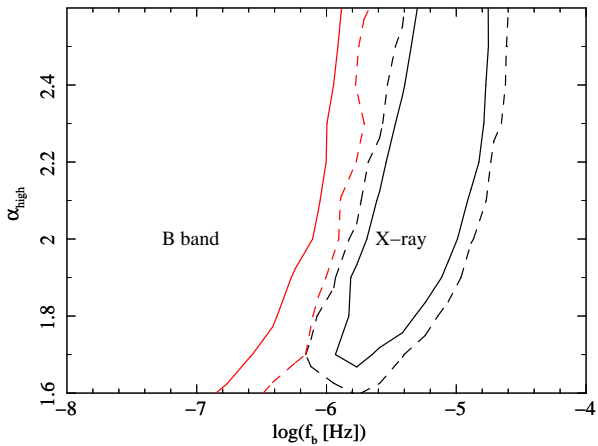}
\caption{Confidence contours for the PDS parameters $\alpha_H$ and $f_b$. The solid and dashed lines represent the 67\% and 90\% probability contours, respectively. The area on the left of the open solid line correspond to the best-fitting region for the B band data and the closed curve on the right contains the best-fitting parameter values for the X-ray data. The parameter spaces allowed for each of the bands do not overlap showing that the PDSs are significantly different.}
\label{contours}
\end{figure}

We also computed the PDS of the B band data, shown by the dotted line in Fig.~\ref{pds}. The long term light curve was used to constrain the PDS at frequencies $3\times10^{-8}-2\times 10^{-7}$ Hz and the intensive light curve covered the range $2\times10^{-7}-7\times 10^{-6}$ Hz. We fixed $\alpha_L=0.8$, i.e.\ the best-fitting value found for the X-ray PDS, for direct comparison with those data. 
Fig.~\ref{contours} shows the 66\% (solid lines) and 90\% (dashed lines) confidence contours for $\alpha_H$ and $f_b$ of the optical and X-ray data. The parameter space regions do not overlap at 90\% significance, showing that the power spectra are significantly different (i.e. they cannot have the same $\alpha_H$ and $f_b$ values simultaneously).  The best-fitting bend frequency is below the lowest frequency probed and the B band PDS is consistent with a single steep power law model.

\section{X-ray/Optical Cross correlation}
\label{ccf}
We calculated the cross correlation between the X-ray and optical light curves using the discrete correlation function (DCF) described in \citet{dcf} and the z-transformed DCF of \citet{alexander}. These cross correlation functions measure the degree of correlation between the two light curves as a function of $\tau$, the displacement of one of the light curves on the time axis. A positive value of the $\tau$ will indicate that X-rays lead the optical fluctuations. 

\begin{figure}
\psfig{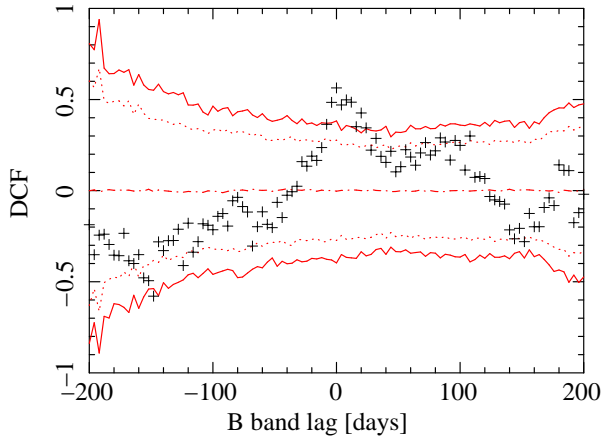}
\caption{The DCF derived using all of the available long timescale X-ray and B-band data shown in Fig.~\ref{lcs} is plotted as black crosses, positive lag values correspond to X-rays leading. The solid lines represent the 99\% extremes of the distribution of simulated X-ray light curves when correlated with the real B band light curve, the dashed lines represent the 95\% extremes of this distribution. The correlation peak around time lag=0 is significant above the 99\% level. The case for X-ray vs V band is identical.}
\label{ccf_B_all}
\end{figure}

We first computed the long term DCF between X-ray and B and V light curves. The intensive sections were resampled by taking 1 point every 4 days in the optical bands and 1 point every two days in the X-rays, to match the sampling of the rest of the corresponding light curve. As the rest of the X-ray light curve was sampled partly every 2 and partly every 4 days, we binned the re-sampled data into 4 day bins to obtain uniform sampling and probe similar time-scales throughout the light curve. We then computed the DCF for each year-long segment and combined the resulting functions, weighting each segment by the number of points in each lag bin. The B and V band light curves are very similar so the DCF between either of them and the X-rays are virtually identical. The long term B vs X-ray DCF is shown in Fig.~\ref{ccf_B_all}. As a separate check we also used the complete light curves (both segments together) resampled and binned into 4-day bins as above and computed the z-DCF. The results were consistent with the DCF results.

We estimated the significance of the DCF by calculating the DCF between the observed optical light curve and 2000 simulated, uncorrelated X-ray light curves. The simulated X-ray light curves were generated following the PDS form and parameters described in Sec.~\ref{psd} using the method of \citet{timmer}, and resampled and rebinned exactly as the real data. This method used to produce the simulated light curves uses random phases for the power spectrum components, while in the real data there might be coupling between these phases, that produces for example the rms-flux relation in the X-ray light curves \citep{rmsflux}. We therefore note that we did not take into account the non-linearity of the light curves when estimating the significance of the DCF, which might have an effect on the significance levels. For each simulation, the DCF was calculated for each of the two segments and then combined, just as was done with the real light curves.  The median, 95\% and 99\% limits of the distribution of simulated DCF values are plotted in Fig.~\ref{ccf_B_all} in dot-dashed, dotted and solid lines, respectively, showing that the central peak measured between the real X-ray and B band light curve is higher than more than 99\% of simulated data. No other peaks or dips in the DCF between the X-ray and either B or V band appeared significant within $\tau \sim 200$ days. We repeated this procedure applying the z-DCF to the complete real and simulated light curves in a consistent way. The central peak in the z-DCF between X-ray and B bands was significant at the 98.4\% level and between X-ray and V bands at the 97.5\% level.

\begin{figure}
\psfig{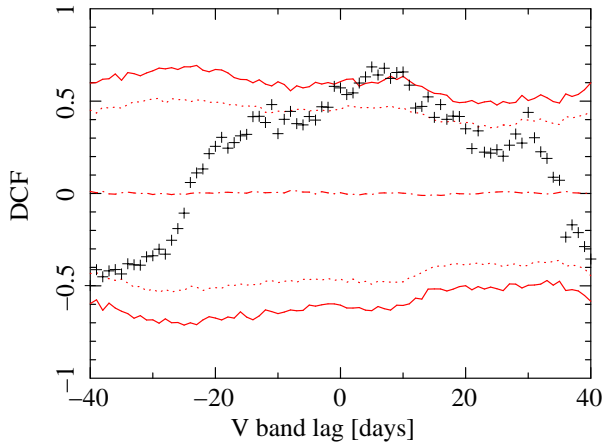}
\caption{DCF derived using only the data from the period of intensive monitoring, shown in Fig.~\ref{intensive} in the X-ray and V bands. We used the daily-sampled V band data together with the 3-times daily sampled X-ray light curved, binned to 1 day bins, to estimate the central peak of the DCF more accurately. The central DCF peak is displaced to positive lags indicating that the short term V band fluctuations lag the X-rays by approximately 5 days. The solid lines represent the 99\% level for the correlation, estimated using simulated X-ray light curves of the appropriate length and sampling. The case for X-ray vs B band is identical.}
\label{ccf_V_int}
\end{figure}

The short time-scale correlation was probed in more detail using the intensive-monitoring light curves shown in Fig.~\ref{intensive}. These include daily sampled optical data for 2.5 months and X-rays sampled three times per day for four months. As shown above in Sec.~\ref{psd}, the X-ray light curve has considerably more variability power than the optical bands on these shorter time-scales. Fortunately, as the sampling of the X-ray band is more intensive we are able to cancel part of the high frequency variability in this band by binning the light curve in 1 day bins. In this way, the correlation between fluctuations on time-scales longer than a day in both bands becomes clearer. We calculated the DCF between the 1-day binned X-rays and the V and B band intensive light curves. The significance of the correlation was estimated as above and the central peak of the DCF was higher than 99\% of the 2000 simulated DCFs for all light curve pairs. The intensive-sampling DCF is shown in Fig.~\ref{ccf_V_int} together with the significance level curves, in this case we show the X-ray vs V band but note that the X-ray vs B band DCF is virtually identical. We note that the X-rays retain much more short time-scale variability power than the optical bands even after binning the X-ray data in 1-day bins. 

Although the light curves are significantly correlated, the CCF peak only reaches values $\sim$0.7, indicating that part of the variability in X-ray and optical bands is incoherent, which makes the delay harder to see directly from the light curves. A first glance at Fig.~\ref{intensive} seems to indicate that the optical bands are leading the X-rays by $\sim 10$ days because of the sharp drop in flux in both bands around the middle of the light curve. The CCF analysis, however, reveals that the correlation at that lag only reaches a value of $\sim 0.5$ and that it is not significant. The only significant peak in the CCF is at positive lag values, with optical bands lagging, which make the overall intensive light curves match better. We note that the significant peak in the CCF at very small lag comes from the low amplitude X-ray/optical fluctuations, as shown in Fig.~\ref{norm_lcs}.

\subsection{Time lags}

Time delays between the different light cures were estimated from the centroid of the central peak in the respective correlation functions. The centroid was calculated as the weighted average of DCF values above 60\% of the DCF maximum. The errors were calculated using the bootstrap method of \cite{bootstrap} by selecting randomly 67\% of the points in each light curve and recording the lag centroid of the resulting DCF. We repeated this procedure 2000 times for each pair of light curves to measure the median and spread of centroid values. The errors quoted correspond to the 67\% limits of the distribution of centroid values on either side of the median.    

The time lags from the long term light curves are $\tau=6.6^{+7.2}_{-6.0}$ days between X-ray and B, $\tau=5.8^{+6.2}_{-3.9}$ between X-ray and V and $\tau=0.3^{+3.5}_{-2.4}$ between B and V. The short, intensively sampled data produced lags of $\tau=5.7^{+2.6}_{-2.5}$ between X-rays and B, $\tau=6.3^{+2.7}_{-3.1}$ between X-rays and V, and $\tau=0.4^{+1.1}_{-1.2}$ between B and V. All the lag measurements are summarised in Table~\ref{T1}, positive lag values indicate higher energy band leading. In all cases the lags from the long and short term light curves are consistent. Both, V and B bands lag the X-rays significantly, from the intensively sampled light curves we measure these lags to be at least 3 and not more than 9 days long. The V and B band fluctuations are simultaneous within our time-resolution, with a delay of V behind B between -0.8 and 1.5 days. 

\begin{table}
\label{T1}
\begin{tabular}{lll}
Energy bands&\multicolumn{2}{c}{Time lag [days]}\\
&long term&intensive\\
\hline
X-ray vs B& $6.6^{+7.2}_{-6.0}$&$5.7^{+2.6}_{-2.5}$\\
X-ray vs V&$5.8^{+6.2}_{-3.9}$&$6.3^{+2.7}_{-3.1}$\\
B vs V&$0.3^{+3.5}_{-2.4}$&$0.4^{+1.1}_{-1.2}$\\
\hline
\end{tabular}
\caption{Time delays in days between different pairs of energy bands, positive values indicate the first band leads. The lags were calculated using long-term, 4-day sampled light curves (middle column) and short-term, daily sampled light curves (right column).}
\end{table}

\begin{figure}
\psfig{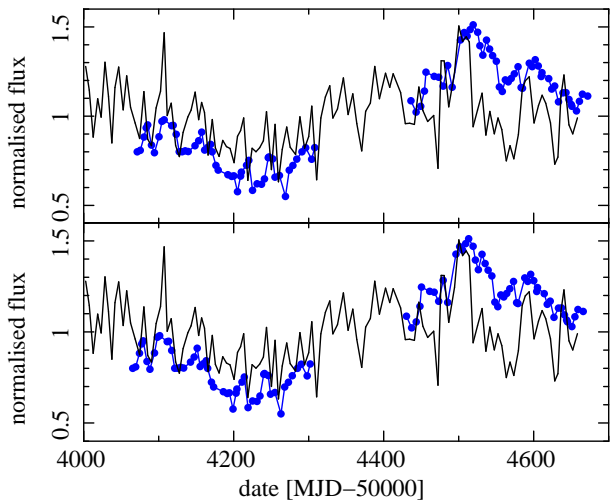}
\caption{X-ray (black lines) and B band (blue dots joined by solid lines) light curves normalised to their mean fluxes. In the bottom panel the B light curve has been shifted back in time by 6 days, i.e. by the time delay indicated in the DCF. The long term trends and short term ($\sim$ 10s of days) peaks are well aligned but the variability amplitudes are different.}
\label{norm_lcs}
\end{figure}

Figure~\ref{norm_lcs} shows the B and X-ray light curves normalised to their mean fluxes, where the B data has been shifted back by 6 days, according to the time lag measured. It is clear that in this way the peaks in both light curves match very well, especially at low optical fluxes. The long term trend has slightly higher amplitude in the B band while the amplitude of the rapid fluctuations is much larger in the X-rays, as expected from the analysis of the PDS. 

\section{Discussion}
\label{discussion}

The measured lag of optical behind X-ray fluctuations suggests a reprocessing scenario where at least part of the optical variability is produced by the variable X-ray heating. In this case, the delay, interpreted as a light travel time, gives the distance between the X-ray corona and the location of the reprocessor.   The black hole mass of \ngc\ is $3\times 10^{7} M_\odot$ \citep{revmap} producing a light crossing time for 1 gravitational radius of $1 R_g/c=150$ s. The time delay of optical behind X-ray fluctuations of $\sim 6\pm 3$ days therefore corresponds to a light travel time through a distance of $\sim 3500\pm1750 R_g$. We note however that the region on the disc where the amount of reprocessed flux peaks will not necessarily correspond to the region of peak intrinsic emission due to internal heating, since X-ray heating may dominate at larger radii as will be discussed below in Sec.~\ref{intrinsicplusreprocessed}. 

\subsection{Transfer functions in the reprocessing scenario}\label{transfersec}
Not all the optical variability can arise from reprocessed X-rays, however, as is evident from the power spectrum analysis. The PDS of the B band data shows that the variability power continues to increase towards lower frequencies, at least down to frequencies $4\times 10^{-8}$ Hz, or correspondingly time-scales of $\sim 300$ days, while the X-ray variability power breaks and flattens for time-scales below 2 days. If all the optical variability was produced by reprocessing, the X-ray fluctuations on time-scales shorter than 100 days would have to be smoothed-out by differential light travel time to opposite sides of the reprocessor, requiring it to have a minimum radius of $\sim50$ light days. In this case, the average delay between optical and X-rays would be of the same order, much longer than the measured $\sim 6$ day lag. 

\begin{figure}
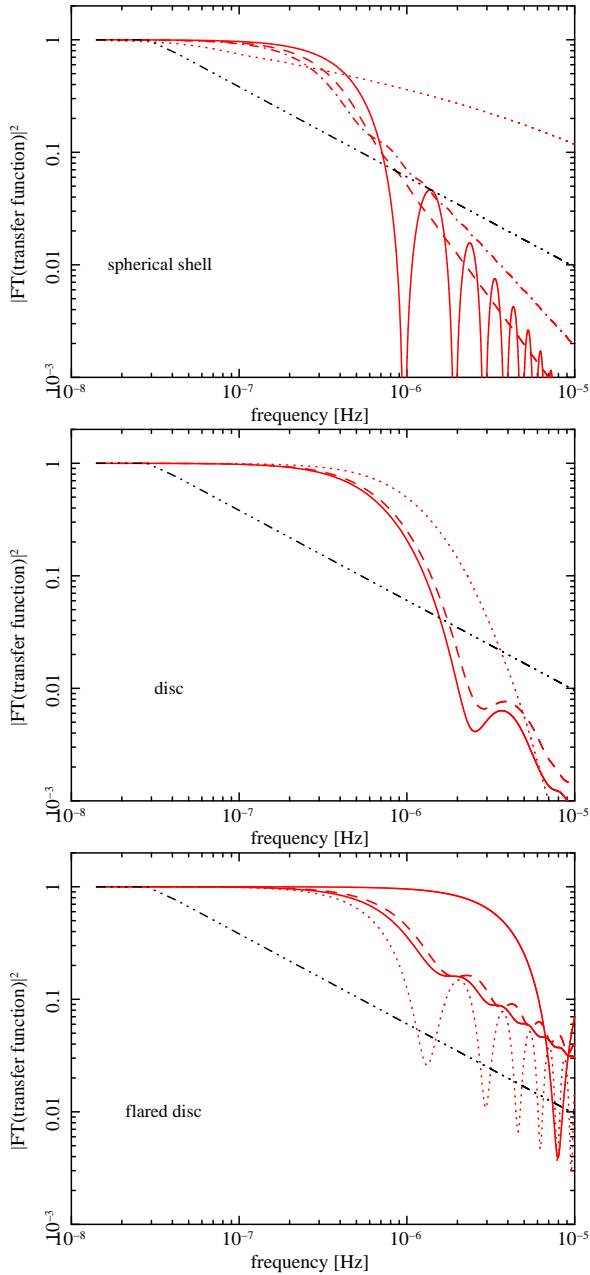

\psfig{file=arevalo_fig9.ps,angle=270,width=8.0cm}
\psfig{file=arevalo_fig10.ps,angle=270,width=8.0cm}
\psfig{file=arevalo_fig11.ps,angle=270,width=8.0cm}
\caption{The ratio between the PDS model fits to the X-ray and B bands from Fig.~4 is shown by the triple-dot-dash lines in all panels. The other curves represent the squared Fourier transform of the transfer functions calculated for different geometries of a reprocessor, as labelled in each panel. For each geometry we calculated transfer functions for different sets of model parameters, with the constraint that the delay between primary and reprocessed light curves should match the observed 6 days (see text for model and parameter descriptions). If the optical variability was entirely produced by reprocessing of the observed X-ray light curve then the PDS ratio would match the Fourier Transform of the transfer function of the correct reprocessor, i.e.\ Eq.~2 would hold. We see that the 6-day delay constraint strongly limits the smoothing power of the transfer function below $\sim 2\times 10^{-7}$ Hz, so none of the reprocessors can smooth out optical variations enough while producing the observed delay. }
\label{transfer}
\end{figure}

To exemplify this argument we constructed transfer functions for different reprocessing geometries.  If the optical variability arises only from X-ray reprocessing then the transfer function has to produce both the smoothing of fluctuations down to time-scales of $\sim 300$ days and a delay of $\sim 6$ days between primary and reprocessed flux. In this case, the variable part of the optical light curve would be $o(t)=x(t)\otimes i(t)$, where $x(t)$ is the X-ray light curve, $i(t)$ is the transfer function and $\otimes$ denotes convolution. The relation of their Fourier transform would therefore be $O(f)=X(f)\times I(f)$, so that  

\begin{equation}
|I(f)|^2=\frac{|O(f)|^2}{|X(f)|^2},
\end{equation}
where $|O(f)|^2$ and $|X(f)|^2$ correspond to the PDS of the optical and X-ray light curves, respectively, as calculated in Sec.~\ref{psd}. The plots in Figure \ref{transfer} compare this PDS ratio (calculated from the fits to the PDS using Eq. 1 and shown here in green triple-dot-dashed lines) to the $|I(f)|^2$ functions predicted for each case. We note that the PDS ratio was calculated for optical and X-ray light curves observed simultaneously, so the variations in the PDS shape expected due to the red noise nature of these light curves is not relevant. 
    
The reprocessor geometries considered were a spherical shell distribution of small clouds, a flat optically thick disc and an optically thick disc which flares out at large radii. These were chosen to mimic structures that probably exist in the vicinity of the black hole such as an optically thick accretion disc and the broad line region (BLR) clouds. In all cases, model parameters were adjusted to produce the observed average delay of 6 days between primary and reprocessed emission, calculated as the expectation value of the transfer function. The spherical shell produces the strongest smoothing of long-term trends for a fixed time lag, compared to the other geometries. This is because there is no delay between the intrinsic emission and the response from the side of the sphere closest to the observer. In disc geometries (except for an edge-on view) the intrinsic flux must first travel to the nearest side of the reprocessor, which is located out of the line of sight, delaying the start of the response. Therefore, the same `smoothing power' can be associated with a longer delay, e.g.\ if the disc is viewed face-on and the reprocessing region is concentrated in a narrow ring then there will be no smoothing of the response but the delay can be arbitrarily long, for an arbitrarily large ring radius. Evidently, the spherical shell is the best bet to smooth out the long-term variability while producing a short lag, as required by the data, but we explore other geometries too for further analysis. We detail each model in the following subsections. 

\subsubsection{Spherical shell}
The first geometry investigated comprised a spherically symmetric distribution of small clouds that capture X-rays from the central source and re-emit a fraction of this flux in the optical bands. Given the symmetry of this setup the only parameters are the radius of the shell $R$ and the polar angle $\theta$ of each position on the shell to the line of sight of the observer. The delay of the response with respect to the primary emission is $\tau(\theta)=R/c\quad (1-\cos{\theta})$.  Assuming that the clouds radiate isotropically and that they do not obscure one-another, the re-emitted flux seen by the observer at a given delay $\tau$ is simply proportional to the area on the shell that corresponds to that delay, i.e. $dA(\tau)/d\tau=2\pi R^2\sin{\theta} d\theta/d\tau =2\pi R^2$, for $\tau$ between 0 and $2R/c$. The $|I(f)|^2$ that corresponds to this setup is shown by the solid line in the top panel of Fig.~\ref{transfer}, where $R$ was fitted to reproduce the observed lag. The 'ripples' in this profile are produced by the sharp drop of the transfer function at $\tau=2R/c$.

To smooth out this drop,  the reprocessing region can span a range of radii rather than being a single thin shell. In this case the total transfer function will be the sum of the contributions from all sub-shells, weighted by the fraction of the sky covered by the clouds on each sub-shell $C(R)$, $i(t)=\sum_{j}C(R_j)A(R_j)$. We used this setup to construct the rest of the functions shown in the top panel of Fig.~\ref{transfer}, taking the thickness of each sub-shell as proportional to its radius.  

The functions shown in the top panel of Fig.~\ref{transfer} correspond to the thickest possible shells for each covering fraction distribution, $C(R)$, using an inner radius of 10 $R_g$ and outer radius fit to match the observed delay. The $|I(f)|^2$ functions changed smoothly from the thin shell profile to the corresponding thickest shell profile for intermediate configurations. The dashed line corresponds to a $C(R)$ that peaked at the middle of the shell and dropped smoothly to zero at the inner and outer boundaries. The dot-dashed line corresponds to a constant $C(R)$, where each sub-shell contributes an equal amount of reprocessed flux. The transfer function in this setup strongly peaks at small lags, where all sub-shells contribute and drops off gently towards larger lags. For very thick shells, as in the case shown in the figure, this sharp peak in the transfer function transfers some variability power on short time-scales, rising the high frequency tail of $|I(f)|^2$. This effect is more extreme in the last case, shown by the dotted line, where the covering fraction drops with radius as $1/r$ so that thick shells make the transfer function even more peaked at small lags and most of the variability power is therefore transferred. We note that this configuration over predicts the short timescale variability without reducing the long term variability enough.

\subsubsection{Truncated flat disc}
The second geometrical setup corresponded to the response of an infinite plane illuminated by an X-ray source at a height of 10$R_g$ following the prescription given in \citet{reprocessing}. We varied the inner truncation radius of the disc in order to fit the lag. The truncation is necessary because otherwise the reprocessed flux peaks at a small radius, making the light travel time to this region, and hence the delay, too short. Truncating the disc further out forces the reprocessing region to appear at a larger distance from the X-ray source and determines the delay. Intrinsic disc flux expected for a standard accretion disc \citep{shakura} was added to the reprocessed flux, for each radius and time delay, to calculate the B band response. The curves shown in the middle plot of Fig.~\ref{transfer} correspond to a face-on view (least smoothing at long time-scales), an intermediate viewing angle of $45^o$ and an edge-on view (most smoothing on long time-scales). The inner truncation radius needed to produce a 6 day lag in each case was 1950, 1550 and 1300 $R_g$, respectively.  

\subsubsection{Flared disc}
The third geometry corresponds to a flared disc, illuminated by a central X-ray source. The region of the disc where the scale height increases presents a large area facing the illuminating source and dominates the reprocessed emission. This scenario can therefore produce a large lag without requiring a large inner truncation radius of the disc. 

For the transfer function calculation, we assumed that the disc surface bends upwards sharply at a radius $R_f$ and the normal to the flared surface makes an inclination angle $a$ with respect to the plane of the flat section of the disc and has length $S$. We neglect differences in emissivity and time delays from different vertical positions on the flare surface. The effective area of this region that receives emission from a central primary source is $dA=(R_f\times S+S^2\sin{a}) \cos{a} d{\theta}$. If $\theta$ is the azimuthal angle on the disc from the projection of the line of sight, the delay from each position is $\tau(\theta)=R_f/c \quad (1-\cos{\theta}\sin{i})$ and the area seen by the observer is $dA'(\tau)/d\tau=(\cos{i}\sin{a}-\sin{i}\cos{a}\cos{\theta})dA/d\theta \quad d\theta/d\tau$, where $i$ is the viewing angle with respect to the axis of symmetry of the disc. When $i>a$ the near side of the disc is seen from behind, which makes $dA'$ negative, as the disc is assumed to be optically thick, this area does not contribute to the transfer function.  

The curves shown in the bottom panel of Fig.~\ref{transfer} correspond to different viewing angles $i$ and opening angles $a$. The strongest smoothing shown is produced for $a=70^o$ and $i=45^o$, requiring a value of $R_f=3150R_g$. The intermediate curves show cases where the opening and inclination angles are similar so the near side of the flared disc is viewed almost edge-on. In these cases the reprocessed flux concentrates only at the far side of the disc, which is viewed more face-on. For this reason the response is retarded so the lag can increase while producing less long-term smoothing. The curves shown correspond to $a=i=75^o$ and $a=i=45^o$ with flare radii of $R_f=2350R_g$ and $R_f=2700 R_g$, respectively. The least amount of long-term smoothing is produced for face-on viewing angles, $i=0$ in this case shown with an opening angle of $a=45^o$ and a radius $R_f=3400R_g$.

\subsection{Intrinsic plus reprocessed optical variability}
\label{intrinsicplusreprocessed}
We note that in all the transfer functions described above, the required lag of 6 days strongly limits the smoothing power below $\sim 2\times 10^{-7}$ Hz so the observed ratio of X-ray versus B band PDS cannot be reconstructed by a single transfer function. This adds to the fact that the B band has slightly larger long-term variations in normalised flux than the X-rays, which is hard to reconcile with a picture where these large amplitude optical trends are produced by reprocessing. We therefore conclude that another source of long term optical variability must exist and it is probably produced by the intrinsic optical emitting region.     

As shown in \cite{MR2251} rapid optical variability, imprinted by reprocessing on a long-term intrinsic fluctuation, can shift the CCF peak to the delay produced by reprocessing, regardless of the delay between the long term fluctuations. This is true even if the long term, intrinsic fluctuations in the optical bands have much larger amplitude than those produced by reprocessing. It is therefore possible that a small amount of the optical flux is produced by reprocessing of X-rays at a large distance from the corona, while the rest of the optical emission and variability are produced intrinsically by the accretion disc at a different radius. In this case, the measured lag is the inprint of X-ray reprocessing on optically thick material at some distance from the X-ray source, which produces the rapid optical variability and 6-day delay. We now estimate the possible location of the optical intrinsic and reprocessed emitting regions.    

A radiatively efficient, optically thick disc radiates approximately as a black body with a radially-dependent characteristic temperature and a total energy output defined by the gravitational energy loss. Assuming such a disc emits the optical flux observed in \ngc\ we can calculate the location of the optical emitting regions. We used the prescription of flux as a function of radius given in \citet{trevesetal88} to calculate the local temperature and optical band flux for a mass of $3\times 10^{7} M_\odot$. We fitted the disc accretion rate in this formula to reproduce the observed optical flux obtaining a value of $\dot m=0.01$ of the Eddington value. We note that this accretion rate is lower than that estimated from the bolometric luminosity because it will only produce the thermal optical/IR emission from the accretion disc. It does not include the fraction of power directed into the corona (which results in X-ray emission) nor the additional IR flux directed into the line of sight by reprocessing structures such as the torus. The resulting 50\% (90\%) of the B band intrinsic disc emission is contained within 73 (208) $R_g$. Given that the long term variability is probably intrinsic to this optical emitting region, we estimate the relevant characteristic time-scales. At a radius of 100$R_g$, which contains 60\% of the B and 50\% of the V flux, the dynamical time-scale (time it takes orbiting material to travel 1 rad of a Keplerian orbit) is 6.7 hours. Optical variability is observed on time-scales of 300 days, i.e.\ $\sim 1000$ times longer than the dynamical timescale at this radius, so accretion rate fluctuations on the local viscous time-scale are not an unreasonable source of this variability. The viscous time-scale of a standard $\alpha$-disc \citep{shakura} is $T_{\rm visc}=T_{\rm dyn}/(\alpha(H/R)^2)$, so in this case $\alpha(H/R)^2=10^{-3}$. Assuming a typical $\alpha=0.1$ this implies a scale height-to-radius ratio $H/R=0.1$, a relatively thick accretion disc. Alternatively, this variability time-scale could correspond to the thermal time-scale of the thin disc, as discussed in e.g.\ \citet{liuht,kelly_optvar}.   

\begin{figure}
\psfig{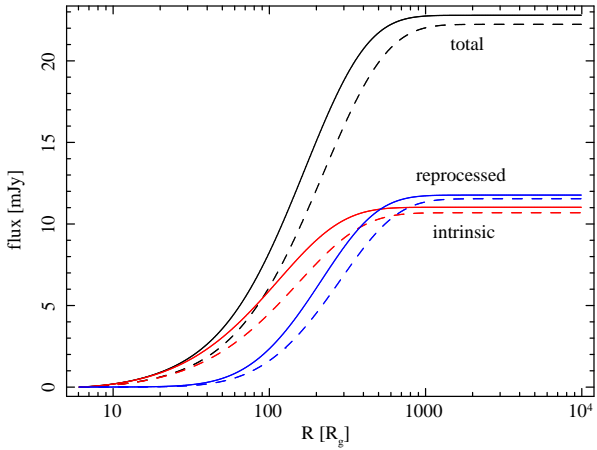}
\caption{Cumulative flux as a function of radius R in the B (solid lines) and V (dashed lines) bands. The intrinsic flux is emitted thermally by an accretion disc while an X-ray source added on the axis of symmetry enhances the disc emission by thermal reprocessing. The curves marked `reprocessed' is the difference between the intrinsic and total curves. The reprocessed flux peaks at a larger radius than the intrinsic emission and it is this larger value that must be used to interpret the delay between X-ray and optical fluctuations. }
\label{profile}
\end{figure}

The location of the \emph{reprocessed} optical emission depends on the location of the X-ray source and the geometry of the reprocessor. We constructed emissivity profiles for the B and V bands assuming a simple geometry of an accretion disc with constant $H/R$, and an X-ray source on the axis of symmetry over the disc. The intrinsic disc accretion rate used was $\dot m=1$\% of the Eddington limit, a BH mass of $3\times 10^{7} M_\odot$ and a value of $H/R=0.1$. The X-ray flux was equal to five times the average 2--10 keV flux measured, to account for X-ray flux outside this band \citep[see][for their X-ray broadband analysis using {\it BeppoSAX} data]{DeRosa} and assumed that 60\% of the impinging X-ray flux were thermalised by the disc \citep[see discussion in][]{MR2251}. Figure~\ref{profile} shows the cumulative optical flux as a function of radius for the intrinsic, reprocessed and total emission for an X-ray source height of 50 $R_g$. The reprocessed curve was calculated as the difference between the intrinsic and total cumulative fluxes. Evidently, the intrinsic and reprocessed emissivities peak at different radii with the reprocessed emission being produced further out. As this is the emission modulated by the variable X-ray flux, we should consider this outer radius to interpret the delay between X-ray and optical fluctuations produced by reprocessing. 

For a face-on view, the parameters used in Fig.~\ref{profile} produce a delay of $\sim 10$ hours for the X-rays to modulate up to 50\% of the reprocessed flux in both B and V bands, too short to explain the observations. Using a thinner disc, down to $H/R=0.001$ does not have a significant effect on this delay, nor does reducing the impinging X-ray flux down to the observed 2--10 keV value (instead of five times this value as was used for the figure). Varying the position of the X-ray source does have a strong effect on the time delay, partly because increasing the height moves the main reprocessing region further out in the disc but mostly because of the longer light travel time down to the disc. The measured delay of a few days would require the X-ray source to be located at $\sim 1000 R_g$ over the disc. 

A different geometry of the reprocessor, however, might also produce the correct time delay. For example a lower X-ray source and a flared disc would present a large area of the disc to the X-ray flux only at large radii, pushing the location of reprocessed emission further out. Alternatively, the reprocessed X-rays might come from a structure similar to the broad line region (BLR). Reverberation mapping campaigns of this source show that the broad H$\beta$ line lags the UV continuum variations by 1--10 days, depending on the line width \citep{revmap}. Therefore, the light travel time from the centre of the accretion disc (probable source of optical/UV continuum) to the BLR is similar to the distance between the X-ray corona and the reprocessing material. In this scenario, the broad line variability tracks the large amplitude optical continuum fluctuations, which modulate the intrinsic optical flux originating close ($\sim 100R_g$) to the black hole. At the same time, the lines can be modulated by the rapid optical fluctuations, produced by X-ray reprocessing at the same radius of the BLR, which already contains a delay to the intrinsic optical and X-ray emission.  
   
Recalling the three geometries of the reprocessor discussed in Sec.~\ref{transfersec} we can now judge their ability to produce the rapid, small-amplitude optical fluctuations only. Although it is not straightforward to disentangle long and short term fluctuations, we can use the low flux section in the first half of the campaign to estimate the amplitude and timescales of what are probably signs of reprocessing. The two big flares on this segment, which show a clear lag to the X-rays, have timescales of 10--20 days and amplitudes of a few mJy in the B band. Also, their relative amplitude is about half or less of the corresponding X-ray flares. The reprocessing mechanism, therefore, should produce a few mJy of optical flux and not smooth out completely the variations on time-scales of tens of days. 

The truncated flat disc reprocessor seems unlikely, on energetics grounds because the X-ray source sees the disc almost edge-on, given the large truncation radius needed for the lag, for any reasonable source height. This small solid angle subtended by the disc produces too little reprocessed flux to make an observable signature in the optical light curves, by many orders of magnitude. This problem is circumvented when using the flared disc because the inclined or vertical portion of the disc can face the X-ray source directly, reaches a higher temperature and produces more optical flux. 

The BLR clouds can have large enough column densities to be optically thick to X-rays \citep[$\sim 10^{24}{\rm cm^{-2}}$,][]{netzerBLR}. Reprocessing in the form of optical emission lines is possible and the effect of their fluctuations on the B and V band fluxes can be estimated from spectral monitoring. The reverberation mapping campaign on \ngc\ presented by \citet{Stirpe}, shows that the lines in this region of the spectrum contribute around 10\% of the B band flux and vary by a maximum ratio of 30\% in flux. Therefore, their contribution to the B and V band variability is too small to produce the observed rapid fluctuations of about 20\%. The BLR, however, can also produce diffuse continuum emission in addition to the lines, as shown by \citet{koristagoad}. These authors calculate the expected BLR contribution to the optical/UV continuum in detail for the case of NGC~5548. They show that 10--20\% of the observed B and V band continua can arise from BLR emission and that the majority of this flux is reprocessed rather than reflected. If a similar amount of continuum emission is contributed by the BLR in \ngc\ and the X-rays (and possibly UV) incident flux is modulating this emission, then the BLR clouds could correspond to the reprocessor that produces the rapid optical fluctuations and the 6-day delay.    

\begin{figure}
\psfig{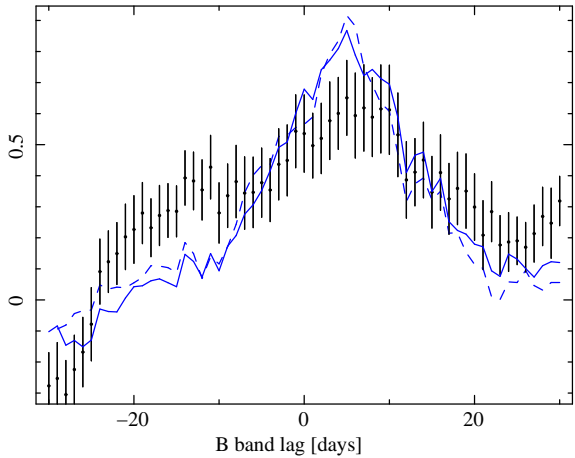}
\caption{DCF between observed B and X-ray bands in markers with error bars. The dashed line represents the DCF between the observed X-ray band and a simulated B band light curve produced by reprocessing the X-ray flux on a flared disc, the solid line corresponds to reprocessing on a spherical shell of clouds. The disc and sphere parameters were fit to match the observed DCF. Both reprocessors can produce the correct delay. The correlation peaks at a higher value for the simulated light curves, probably due to the additional sources of variability in the real optical light curve.}
\label{ccf_bestfit}
\end{figure}

We calculated `reprocessed' B band light curves by convolving the observed X-rays with the transfer function of the flared disc and the spherical shell of Sec.~\ref{transfersec} as these structures are the most promising candidates for the reprocessor. We used the intensively sampled X-ray data to produce the simulated B band light curves and compare the resulting optical vs X-ray DCF with the DCF of the real data, to select the best fit in the $\tau=\pm 30$ day range. In the case of the flared disc we calculated B band emission assuming that the flared portion of the disc emits as a black body at the temperature corresponding to the reprocessed X-ray flux (as the radius is large, intrinsic heating at in this region is negligible). As an additional constraint in the fitting routine we required an average reprocessed flux of at least 1 mJy at the observer location, to reproduce the small amplitude optical fluctuations, which contain the 6-day delay. The height of the X-ray source was fixed at 10$R_g$ and the 2--10 keV flux was amplified by a factor of 5 as above, to include X-ray flux outside of this band. This setup could produce the required amount of B band flux and the correct  delay for a range of parameter values. The best-fitting values were $a=60^o$, $i=45^o$, $R_f=2200 R_g$ and a length of the flared region of $S=600R_g$. In the case of the BLR reprocessor we only calculated the shape of the reprocessed light curve and relied on the numbers given by \citet{koristagoad} to produce sufficient reprocessed optical flux. The best-fitting geometrical parameters for this case where $R_{\rm in}=1000 R_{\rm g}$ and a thickness of four times this radius. The fitted portion of the intensive DCF for both models is shown in Fig.~\ref{ccf_bestfit}. The model parameters are not well constrained in either case and the quality of the fits are similar so this analysis only shows that both structures are feasible reprocessors. 

The simulated reprocessed light curves lag the X-rays by approximately the same time as the real optical data, but their DCFs peak at a higher value than the DCF of the real data. We note that the simulated B band light curves are produced only through X-ray reprocessing, so a good correlation is expected. On the other hand, the real optical light curve probably contains other sources variability, producing the long term trends as argued above, which will lower the correlation peak. 

 \section{Conclusions}
\label{conclusion}

We have monitored the Seyfert galaxy \ngc\ over two years, simultaneously in the X-ray 2--10 keV and optical B and V bands. The flux in all bands is highly variable on all the time-scales probed. Our main conclusions drawn from the variability properties are detailed below.

\begin{enumerate}
\item We recalculated the power spectrum of \ngc\ in the X-ray band. Our new, intensively sampled light curve was programmed to cover the time-scales where a possible QPO had been detected with earlier data \citep{summons}. The high quality new data, however, shows that the entire PDS is consistent with a single-bend power law and the possible QPO is not confirmed. The PDS parameters we found are consistent with the single-bend power law model fit of \citet{summons}, with values of $\alpha_L=0.8, \alpha_H=2.2, f_b=5.8\times10^{-6}$ Hz.

\item B and V bands vary simultaneously, with a delay of no more than 1.5 days. The light curves are almost identical but the fractional amplitude of variations is slightly larger in the B band. These results are consistent with those obtained by \citet{Reichert} who monitored \ngc\ in the UV and optical bands, finding the largest variability at the shortest wavelengths.

\item The fluctuations in optical and X-ray bands are significantly correlated. The optical bands show a delay with respect to the X-rays of $\sim 6\pm 3$ days. Similar values are obtained for both B and V bands and for long-term and intensively sampled light curves.   

\item The PDS of the optical bands is consistent with a single power law of slope $\alpha=1.6$, where the variability power increases continuously toward the longest time-scale measured, $\sim 300$ days. Comparison with the PDS of the X-ray band shows that if all optical variability arises from reprocessing of X--rays, the reprocessor would have to be $\sim 50$ light days across, to smooth out fluctuations on shorter time-scales. This size is inconsistent with the small value of the delay found between optical and X-ray fluctuations (6 days) so another source of optical variability must exist.

\item We interpret the optical variability as arising from both reprocessing and intrinsic accretion rate fluctuations. Accretion rate variations on time-scales of hundreds of days can cause the large amplitude optical fluctuations and propagate to the X-ray emitting corona producing the long term correlation. Reprocessing of X-rays can produce the short-term, small amplitude optical variations explaining the measured positive delay of optical with respect to X-rays. The most likely reprocessor geometries we investigated were an optically thick accretion disc which flares out at large radii and a spherical shell distribution of clouds. In the disc case the scale height increases at a radius similar to the location of the BLR, while in the second case the reprocessor could correspond to the BLR itself if the clouds can produce sufficient optical continua from reprocessed X-rays and UV photons, as discussed in \citet{koristagoad}.      
 
\end{enumerate}

\section*{Acknowledgments} This work has made use of observations
obtained with \xte\ and the SMARTS Consortium telescopes. We thank
H. Nezter for useful discussion. The authors acknowledge financial
support from the following: PA from a Fellowship for International
Young Researchers, Chinese Academy of Sciences and from the
Max-Planck-Institut f\"ur Astrophysik, Garching, PU from an STFC
Advanced Fellowship, PL from FONDECYT grant number 1080603, EB from a
Stobie-SALT Scholarship and IMcH acknowledges support from STFC under
rolling grant PP/D001013/1.

\bibliographystyle{mn2e}
\label{lastpage}

\end{document}